\documentclass[fleqn,10pt]{olplainarticle}

\usepackage{graphicx}
\usepackage{caption}
\usepackage{subcaption}
\usepackage{array}
\usepackage{underscore}
\usepackage{multirow}
\usepackage{subcaption}
\usepackage{gensymb}
\usepackage{adjustbox}
\usepackage{hyperref}
\usepackage{lineno}
\usepackage{setspace}
\usepackage[numbers, square]{natbib}

\renewcommand{\cite}[1]{\citet{#1}}
\title{VascX Models: Model Ensembles for Retinal Vascular Analysis from Color Fundus Images}

\author[1,2]{Jose Vargas Quiros, None}
\author[1,2]{Bart Liefers, None}
\author[1,2]{Karin van Garderen, None}
\author[1,2]{Jeroen Vermeulen, None}
\author[1,2]{Eyened Reading Center, None}
\author[1,5,6,7]{Sinergia Consortium, None}
\author[1,2,3,4]{Caroline Klaver, None}
\affil[1]{Department of Ophthalmology, Erasmus University Medical Center, Rotterdam, the Netherlands}
\affil[2]{Department of Epidemiology, Erasmus University Medical Center, Rotterdam, the Netherlands}
\affil[3]{Department of Ophthalmology, Radboud University Medical Center, Nijmegen, the Netherlands}
\affil[4]{Institute of Molecular and Clinical Ophthalmology, University of Basel, Switzerland}
\affil[5]{Department of Computational Biology, University of Lausanne, Lausanne, Switzerland}
\affil[6]{Jules Gonin Eye Hospital (HOP), Fondation Asile des aveugles, Lausanne, Switzerland}
\affil[7]{Department of Clinical Neuroscience, Laussane University Hospital, Laussane, Switzerland}


\begin{abstract}

We introduce VascX models, a comprehensive set of model ensembles for analyzing retinal vasculature from color fundus images (CFIs). Annotated CFIs were aggregated from public datasets . Additional CFIs, mainly from the population-based Rotterdam Study were annotated by graders for arteries and veins at pixel level, resulting in a dataset diverse in patient demographics and imaging conditions. VascX models demonstrated superior segmentation performance across datasets, image quality levels, and anatomic regions when compared to existing, publicly available models, likely due to the increased size and variety of our training set. Important improvements were observed in artery-vein and disc segmentation performance, particularly in segmentations of these structures on CFIs of intermediate quality, common in large cohorts and clinical datasets. Importantly, these improvements translated into significantly more accurate vascular features when we compared features extracted from VascX segmentation masks with features extracted from segmentation masks generated by previous models. With VascX models we provide a robust, ready-to-use set of model ensembles and inference code aimed at simplifying the implementation and enhancing the quality of automated retinal vasculature analyses. The precise vessel parameters generated by the model can serve as starting points for the identification of disease patterns in and outside of the eye.
\end{abstract}

\begin{document}

\flushbottom
\thispagestyle{empty}
\maketitle

\section*{Introduction}

Interest in measuring retinal calibers dates back to the 1970s when Parr and Spears published their approach to approximate the caliber of the central retinal artery and vein using the calibers of their largest branches \citep{parr1974general}. Since the late 1990s, various semi-automatic methods have been developed to perform these calculations with different levels of human and machine involvement, enhancing the efficiency of such measurements \citep{hubbard1999methods}. Among these developments were prominent tools such as SIVA (Singapore I Vessel Assessment) \citep{wong2006retinal, klein2012relationship, iwase2017new, wei2016retinal, dervenis2019factors}, IVAN \citep{cheung2011quantitative,cheung2010new}, and VAMPIRE (Vascular Assessment and Measurement Platform for Images of the Retina) \citep{perez2011vampire, trucco2013novel, trucco2015morphometric}.

Today the understanding of retinal vasculature and its association with diseases has expanded substantially. A significant focus has been placed on biomarkers such as retinal calibers, tortuosity, and branching characteristics. These biomarkers are now studied not only in relation to ocular diseases but also to systemic conditions, particularly cardiovascular diseases \citep{wong2002retinal, liew2008retinal, cvd-ai}. More recently, attention has turned to the retina as a window into brain pathology, with studies linking retinal vascular changes to neurodegenerative diseases such as Alzheimer's disease \citep{frost2013retinal}. Previous work has shown that automatically extracted vascular features improve the performance of diabetic retinopathy (DR) detection and classification \citep{av-helps}.

Parallel to the scientific advances, deep learning brought about ground-breaking improvements in segmentation quality across a variety of tasks \citep{nnunet}. This triggered the development of multiple datasets, deep learning architectures and models for the segmentation of the main anatomical structures visible in CFIs: vessel segmentation \citep{iternet, vessel-transf}, artery-vein segmentation \citep{lwnet,lunet}, and disc segmentation. The task of assessing fundus image quality has also received attention  \cite{eyeq,qnet,iq-keane}. More recent are efforts to integrate these models into pipelines for fully automated, explainable, vascular feature extraction. Automorph is the most notable example of an open source deep-learning-based pipeline integrating vessel segmentation, artery-vein segmentation, disc segmentation, and vascular feature extraction \cite{automorph}. Automorph has been applied in various studies including on the reproducibility of its features \citep{automorph-app1,automorph-app2,samuel-tortuosity,fd-repro}. The availability of disc segmentation models in Automorph allows for region-based analysis which isolate features from specific areas of the retina. 

Despite these developments, the robust segmentation of retinal morphology (arteries, veins, optic disc, fovea) from CFIs across a variety of patient and imaging conditions remains unsolved. Most publicly available training datasets for artery-vein segmentation are small and homogeneous, often limited to images from a single device, centered on a single anatomic region (fovea or optic disc), and with a fixed field of view. Until recently \citep{leuven}, the largest publicly available dataset for artery-vein segmentation contained only 45 images \citep{hrf}. Additionally, the pathology and demographics of the study population are often restricted, which significantly impacts the generalizability of the models trained on these datasets. This affects the downstream extraction of explainable vascular features, which depend on the quality of such segmentations \citep{samuel-automorph-repro, fd-repro}. This is particularly problematic in analyses of large cohort studies and clinical data, which may encompass a diverse range of devices and capture conditions. To overcome these drawbacks, larger and more diverse training datasets are needed to develop models that work reliably across a broader spectrum of conditions.

In this paper, we present contributions towards a robust analysis of the retinal vasculature through improved segmentation and localization models. Our contributions are the following:

\begin{itemize}
    \item We present models for optic disc, vessel, and artery-vein segmentation and fovea localization, fundamental models in the development of CFI analysis systems, especially of the retinal vasculature. We built our models using public datasets and new annotations by professional graders on diverse sets of CFIs from Dutch studies, mainly the Rotterdam Study.
    \item We benchmark our models against publicly available models \citep{automorph, lwnet} and show significant improvements in segmentation performance. We characterize the performance of the models with respect to image quality as measured using our quality assessment model.
    \item We evaluate the quality of features extracted from model output by comparing them to features extracted from segmentations made by experienced graders.
    \item We publicly release and make open-source our pre-processing and inference code, including model weights.
\end{itemize}





\section*{Methods}

\subsection*{Datasets}

We trained and evaluated our models on a combination of public datasets and images from the Rotterdam Study (RS) and other Dutch studies.

\subsubsection*{Public Datasets}
We collected publicly available datasets from previous work. These datasets have different patient demographics, pathologies, retinal regions (fovea and disc centered), field of view and countries of origin. Table \ref{tab:datasets} shows an overview of these datasets. For complimentary details including capture devices and resolution see Supplementary Table \ref{tab:datasets_appendix}.

\begin{table}[p!]
    \centering
    \captionsetup{justification=centering}
    
    \caption{Overview of datasets for various segmentation and localization tasks. Pathologies: H: healthy, GC - glaucoma (all types), NTG: normal tension glaucoma, HTG: high tension glaucoma, HT: hypertension, DR: diabetic retinopathy, AMD: age-related macular degeneration, PM: pathological myopia, O: other eye diseases. *: 379 of 1200 samples were discarded due to containing no disc segmentation for partially visible discs. $\dagger$: 21 of 1200 CFIs were discarded due to having no associated annotation. }

    \begin{adjustbox}{max width=\textwidth}
    \begin{tabular}{>{\arraybackslash}p{3cm}>{\arraybackslash}p{1cm}>{\arraybackslash}p{1cm}>{\arraybackslash}p{2cm}>{\arraybackslash}p{3.5cm}>{\arraybackslash}p{1.8cm}>{\centering\arraybackslash}p{1cm}}
        \toprule
        \textbf{Dataset} & \textbf{N} & \textbf{Origin} & \textbf{Ages} & \textbf{Pathology} & \textbf{Region} &\textbf{FOV}  \\
        \vspace{1pt} \\
        \multicolumn{7}{c}{\textbf{Vessel Segmentation}} \\
        \midrule
        \noalign{\vskip 5pt} 

        Chase DB \citep{CHASEDB1} & 28 & UK & Children & - & OD & 35\degree  \\
        DRHAGIS \citep{drhagis} & 40 & UK & NR & 10 GC, 10 HT, 10 DR, 10 AMD & M & 45\degree \\
        HRF \citep{hrf} & 45 & GER & NR & 15 H, 15 DR, 15 GC & M & 45\degree  \\
        RETA \citep{reta} & 54 & IND & NR & Signs of DR & M & 50\degree \\
        FIVES \citep{fives} & 800 & CN & 4-83 & H, GC, DR, AMD & M & 50\degree \\
        Leuven-Haifa \citep{leuven} & 240 & BE & 18-90 & 75 NTG, 63 HTG, 56H, 30 O & OD & 30\degree \\
        
        Rotterdam (ours) & 352 & NL & 40+ & \multicolumn{3}{c}{See Figure 1.} \\
        \noalign{\vskip 5pt}
        \textbf{Total} & \textbf{1559} & & \\
        
        \vspace{1pt} \\
        \multicolumn{7}{c}{\textbf{Artery Vein Segmentation}} \\
        \midrule
        \noalign{\vskip 5pt}
        
        RITE \citep{avdrive} & 40 & NL & 25-90 & 7 DR, 33 no DR & M & 45\degree  \\
        HRF-AV \citep{hrfav} & 45 & GER & NR & 15 H, 15 DR, 15 GC & M & 45\degree \\
        Les-AV \citep{lesav} & 22 & NR & NR & NR & OD & 30\degree \\
        Leuven-Haifa \citep{leuven} & 240 & BE & 18-90 & 75 NTG, 63 HTG, 56H, 30 O & OD & 30\degree \\
        Rotterdam (ours) & 215 & NL & 40+ & \multicolumn{3}{c}{See Figure 1.} \\
        \noalign{\vskip 5pt}
        \textbf{Total} & \textbf{562} & & \\

        \vspace{1pt} \\
        \multicolumn{7}{c}{\textbf{Disc Segmentation}} \\
        \midrule
        \noalign{\vskip 5pt}

        ORIGA \citep{origa} & 650 & SGP & 40-80 & Multiple & OD & NR \\
        PAPILA \citep{papila} & 488 & SPN & 15-90 & 87 GC, 333 H, 68 O & OD & 30 \degree  \\
        IDRiD \citep{idrid} & 81& IND & NR & Signs of DR & M & 50\degree  \\
        ADAM \citep{adam} & 821* & CN & $53.2 \pm 15.6$ & 267 AMD, 933 O & M, OD, MP & 45 \degree \\ 
        PALM \citep{palm} & 1179$\dagger$ & CN & $37.5\pm15.91$ & 637 PM, 563 O & M, OD, MP & 45 \degree \\ 
        REFUGE2 \citep{refuge2} & 2000 & CN & NR & 280 GC, 1720 H/O & M, OD, MP & 45 \degree  \\
        Rotterdam (ours) & 1225 & NL & 40+ & \multicolumn{3}{c}{See Figure 1.}  \\
        \noalign{\vskip 5pt}
        \textbf{Total} & \textbf{7464} & & \\

        \vspace{1pt} \\
        \multicolumn{7}{c}{\textbf{Fovea Localization}} \\
        \midrule
        \noalign{\vskip 5pt}
        
        IDRiD \citep{idrid} & 516 & IND & NR & Signs of DR & M & 50\degree  \\
        ADAM \citep{adam} & 1200 & CN & $53.2 \pm 15.6$ & 267 AMD, 933 O & M, OD, MP & 45 \degree \\
        PALM \citep{palm} & 1200 & CN & $37.5\pm15.91$ & 637 PM, 563 O & M, OD, MP & 45 \degree \\
        REFUGE2 \citep{refuge2} & 2000 & CN & NR & 280 GC, 1720 H/O & M, OD, MP & 45 \degree  \\
        
        Rotterdam (ours) & 10908 & NL & 40+ & \multicolumn{3}{c}{See Figure 1.}  \\
        \noalign{\vskip 5pt}
        \textbf{Total} & \textbf{15824} & & \\

        \vspace{1pt} \\
        \multicolumn{4}{c}{\textbf{CFI Quality Estimation}} \\
        \midrule
        \noalign{\vskip 5pt}

        EyeQ \citep{eyeq} & 28,792 & &  \\
        \bottomrule
    \end{tabular}
    \end{adjustbox}
    \label{tab:datasets}
\end{table}

We included several well-known vessel and artery-vein segmentation datasets with diabetic retinopathy (DR) patients \citep{drhagis, hrf, reta, hrfav}. We excluded the DRIVE dataset \citep{drive, avdrive} due to its low resolution ($584\times565$ pixels). We included the larger and more recent FIVES \citep{fives} and Leuven-Haifa \citep{leuven} datasets, which contain a mixture of diseased and healthy patients. On the other hand, most datasets with disc segmentations were originally collected for the assessment of glaucoma \cite{origa,papila,refuge2}. The REFUGE2 dataset, an extension of the original REFUGE dataset \cite{refuge1} is notable for its size (2000 CFIs), various anatomic regions captured, and for containing fovea location annotations. We also included ADAM (age-related macular degeneration) \citep{adam}, PALM (pathological myopia) \citep{palm} and IDRiD (diabetic retinopathy) \citep{idrid}; which also contain fovea positions. 

Finally, we used the Eye-Quality (EyeQ) dataset (Table \ref{tab:datasets}) for the quality estimation algorithm. This dataset contains manual gradings of fundus quality from 28,792 images using a 3-class system. CFIs were graded as one of: unusable/bad, usable and good quality. It is unique for its size and for containing a variety of capture devices and image conditions. The field-of-view of the images in our training set is between 30\degree and 50\degree, which includes the vast majority of standard fundus imaging devices.

\subsubsection*{Rotterdam Datasets}
We complimented the CFIs from the public data sets with images available to the Department of Ophthalmology at Erasmus Medical Center in Rotterdam; mainly the Rotterdam Study (RS). The RS is a prospective population-based cohort study of people living in Ommoord, a district of the city of Rotterdam (~\cite{Ikram2020}). The RS consists of four cohorts, all of which were used in this work. The minimum age of study participants varies between $>55$ in the first cohort and $>40$ in the fourth. Each cohort was followed for multiple rounds of follow-up examinations every 4 to 5 years. Most of the visits in the RS involved the capture of CFIs on both eyes. Due to the multi-decade span of the RS, multiple devices, capture conditions and fields (macula and disc centered) are present in the dataset. 

In addition to the RS, we included images captured during the Randomized Controlled Trial for Age-Related Macular Degeneration (AMD-Life) \citep{amdlife}; which consisted of 150 patients aged 55–85 years imaged with a Topcon 3D OCT 2000-plus and Topcon TRC-50EX  cameras. Images from the Dutch Myopia Study (MYST) \citep{myst} which included younger subjects (mean (SD) age of 47.3 (12.9) years) were also used. These images were captured with a Topcon TRC-50EX.

\subsubsection*{CFI Selection and Weighing}

We used most public datasets in their entirety without filtering, after harmonizing annotations when necessary. The main exception were disc segmentations in the ADAM dataset, where 379 of 1200 samples were discarded due to containing no disc segmentation for partially visible discs. 

Regarding the Rotterdam sets, we sampled CFIs at random from all available images. We included fundus images captured by OCT machines to improve diversity of devices in the dataset. For the vessel and A/V segmentation datasets (Table \ref{tab:datasets}), due to the much larger size of the Rotterdam Study compared to AMD-Life and MYST, we over-sampled CFIs from these sets to make up 15\% of the samples. We used the quality estimation model in Automorph \citep{automorph} to assess image quality and excluded images classified as \textit{unusable}. Additionally, we asked graders to exclude images for which they found the quality to be unusable for the task at hand. 

For disc segmentation, due to the large size of publicly available datasets (Table \ref{tab:datasets}), we first used a model trained on publicly available datasets to automatically identify failure or challenging CFIs. We selected CFIs for which no disc was detected or a disconnected disc mask was produced by the model. A trained grader reviewed these images and identified the ones where the disc was present but incorrectly segmented, and these were included in the training set. In practice, these included many analog, macula-centered images from the Rotterdam sets, which specially challenged the model.

For fovea detection, we used CFIs from a sub-cohort of the RS previously graded for signs of Age-related Macular Degeneration (AMD); which included localization of the fovea for ETDRS grid placement.

\begin{figure*}[ht]
    \centering
    \includegraphics[width=0.85\textwidth]{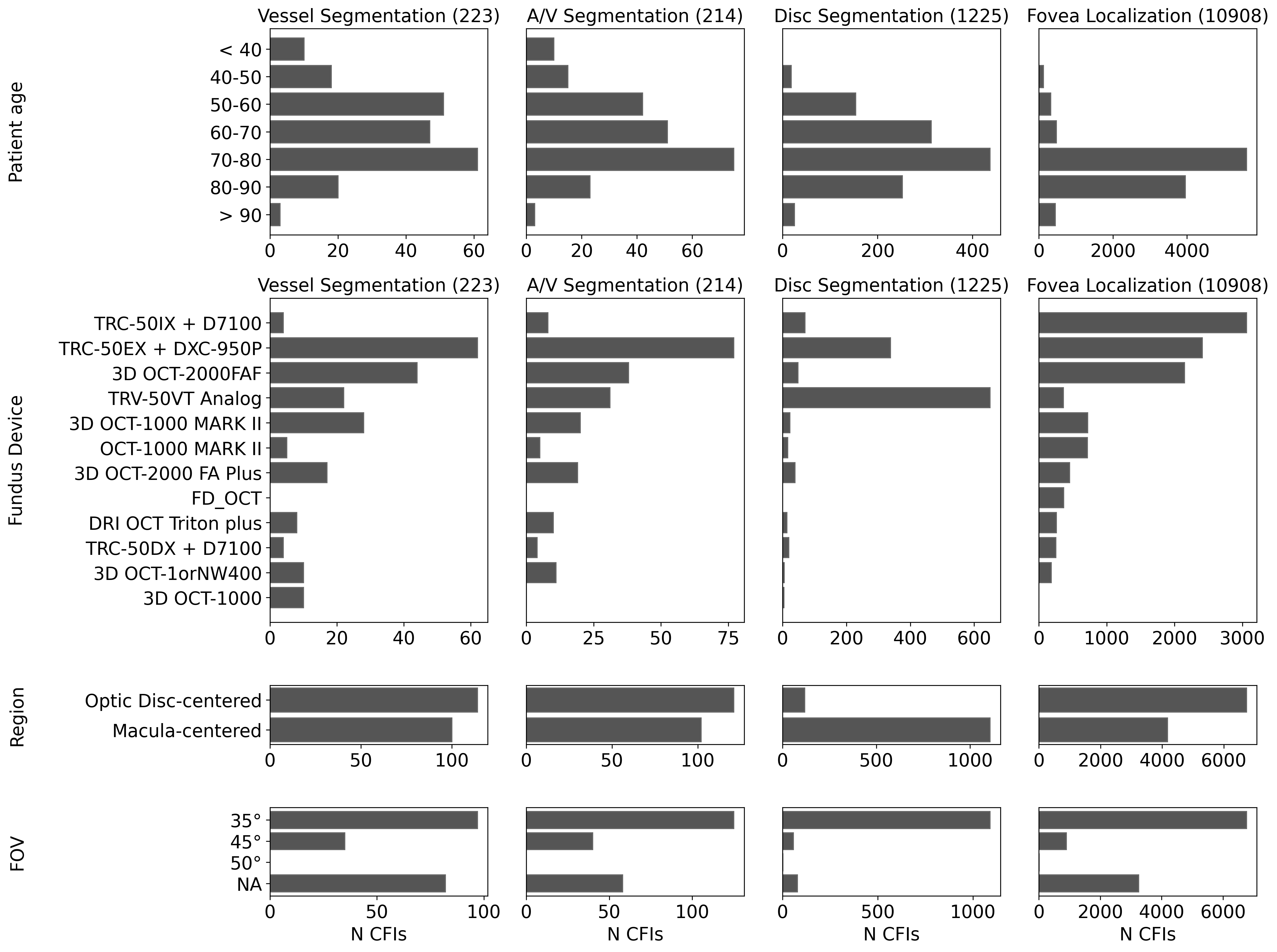}
    \caption{Statistics of the Rotterdam development sets used to train and evaluate VascX models. NA indicates the information was not available.}
    \label{fig:ds_stats}
\end{figure*}

Figure \ref{fig:ds_stats} shows the statistics of the final Rotterdam development sets. Vessel and artery-vein segmentation sets are similar (partly due to overlap) and are closer to the overall distribution of the RS. The disc segmentation dataset, on the other hand, contains more images from the analog devices, which were often the hardest to segment partly due to unconventional disc placement. 

\subsection*{CFI annotation}

\begin{figure}[ht]
    \centering
    \begin{subfigure}[b]{0.49\textwidth}
        \centering
        \includegraphics[height=4.25cm]{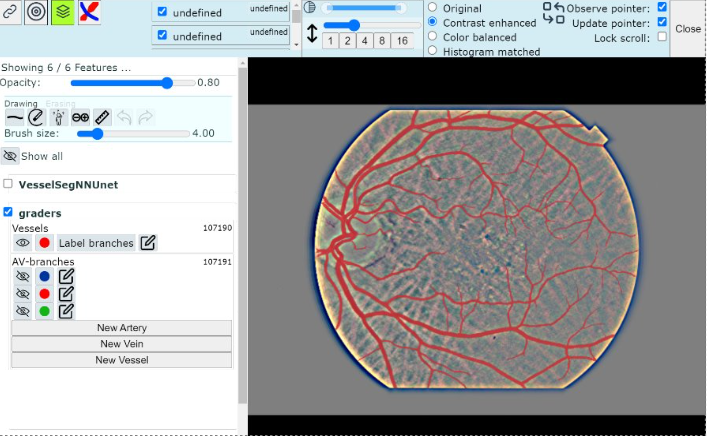}
        \caption{Vessel segmentation: graders corrected binary masks generated by AI, using a digitally contrast-enhanced image to visualize vessels.}
        \label{fig:viewer1}
    \end{subfigure}
    \hfill
    \begin{subfigure}[b]{0.49\textwidth}
        \centering
        \includegraphics[height=4.25cm]{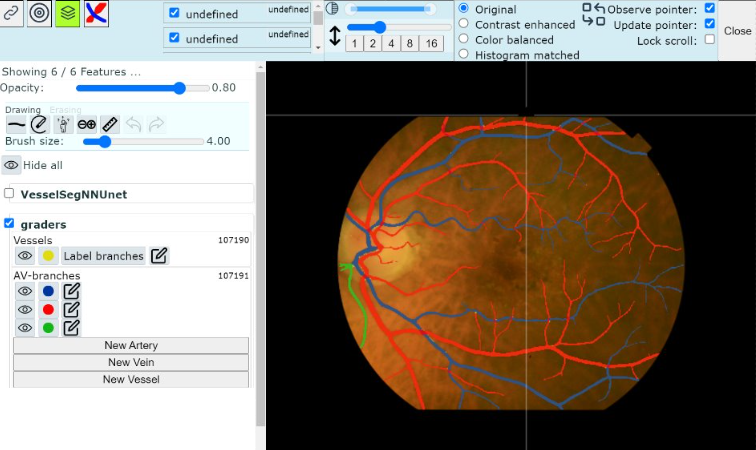}
        \caption{Artery-vein segmentation: graders colored the vessel masks into two separate (overlapping) masks for arteries and veins using a custom interface.}
        \label{fig:viewer2}
    \end{subfigure}
    \caption{Screen capture of the software used for all the annotations on Rotterdam CFIs.}
    \label{fig:viewer}
\end{figure}

Images were annotated using custom software for ophthalmological image annotation \citep{viewer}. Four professional graders participated in the process. The graders worked on desktop computers with a drawing tablet for segmentation. The annotation process was different per annotation type:

\begin{description}
    \item[Vessel segmentation] We followed an AI-assisted process by initializing the drawing interface with with outputs from vessel segmentation models. Initially (first 200 images) we made use of models trained on public data (including Automorph \cite{automorph}) to later switch to a model trained on both public data and annotations from previous iterations of this process. The graders were instructed to independently fix the masks produced by the model, including missing vessels and over-segmentation. A digitally enhanced version of the image with increased contrast of vessel edges was used to aid the graders (Figure \ref{fig:viewer1}).

    \item[Artery-vein segmentation] We resolved connectivity issues at artery-vein crossings by annotating arteries and veins on separate layers. An interface was developed specifically for artery-vein segmentation (Figure \ref{fig:viewer2}) to allow the graders to start from a vessel segmentation (no A/V distinction) and color arteries and veins into two independent masks or layers using drawing tools. A third layer - \textit{Unknown} was added for vessels that could not be recognized as either. They were asked to color A/V crossings on both layers (overlapping). They were able to visualize and correct each layer independently, or all the same time. Another feature allowed them to visualize the connected components of each mask in different colors to easily find mistakes in connectivity. The entire process consisted in 1) correcting mistakes in the provided mask (vessel segmentation); 2) coloring the mask into A/V and unknown colors/layers; 3) Verifying connected components on the A/V masks and filling any gaps.

    \item[Disc segmentation] A single professional grader performed this annotation using the same drawing interface used for binary vessel segmentation (Figure \ref{fig:viewer1}). The grader colored the entire optic disc area excluding any peri-papillary atrophy (PPA) region to match the annotations on public datasets \citep{g1020, origa, refuge2}. In cases of doubt due to poor image quality or unclear disc boundaries, a consensus was reached between the three graders. Cases where the image quality was too poor to reach consensus were excluded.

    \item[Fovea localization] The fovea was localized by four graders who followed a standard grading protocol which included placing an ETDRS grid; overlaid on the images.
    
\end{description}

\subsection*{Model Training}

We implemented an improved CFI bounds detection and pre-processing procedure and used well-known model architectures such as U-Net for model training.

\subsubsection*{Data Pre-processing}

\begin{figure}[!ht]
    \centering
    \includegraphics[width=\textwidth]{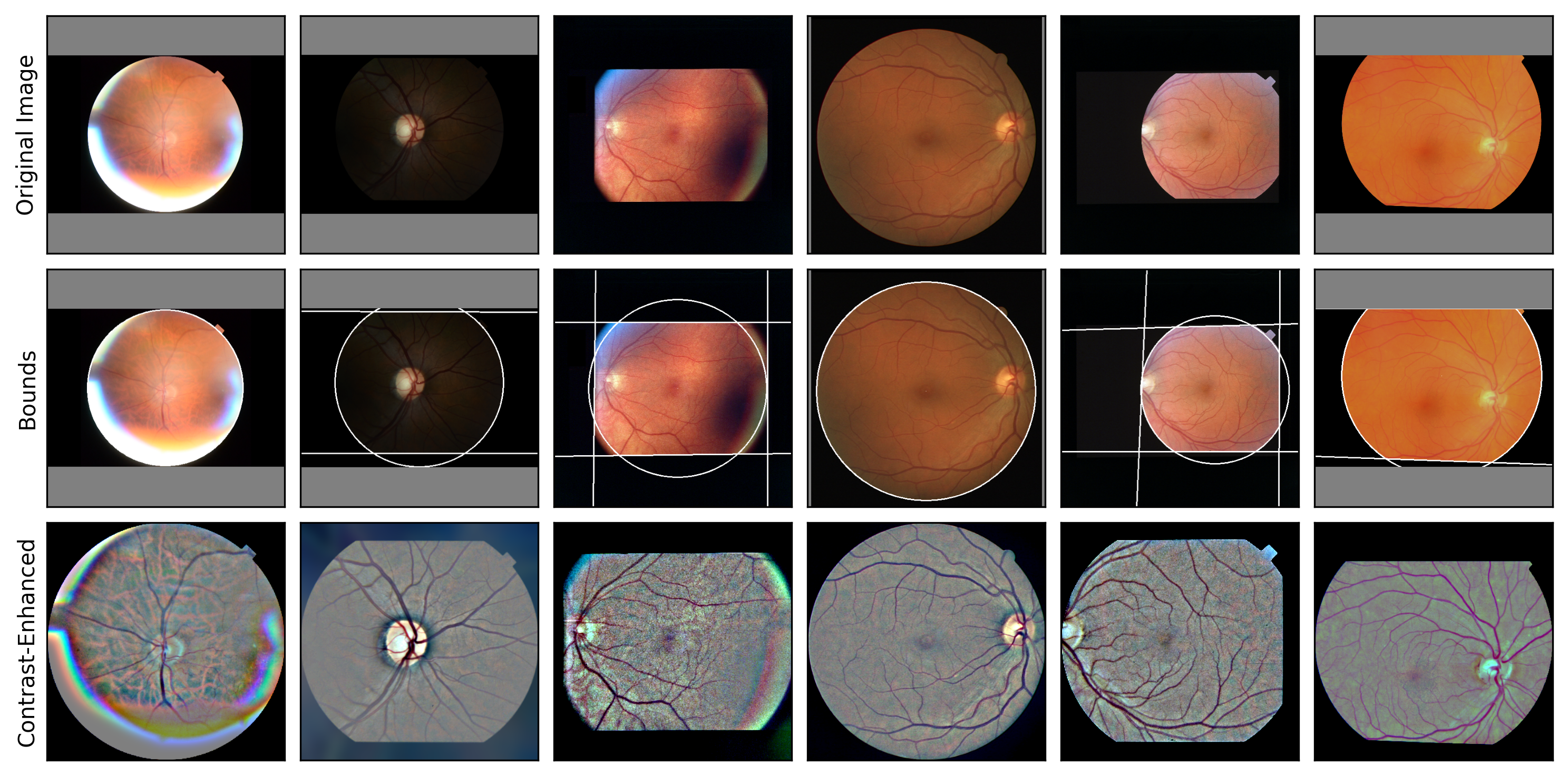}
    \caption{Sample of pre-processed images from the Rotterdam sets, showing: original images (first row), the results of bounds detection defined by the intersection of a circle and optional lines (second row) and the cropped and contrast-enhanced image (third row).}
    \label{fig:prep}
\end{figure}

For all models, training and testing CFIs were pre-processed via:

\begin{enumerate}
    \item Detection of the CFI boundaries defined by a circle and top, bottom, left and right lines (see Figure \ref{fig:prep}).
    \item Cropping of the CFI along its boundaries into a square image and resizing to $1024x1024px$.
    \item Contrast enhancement via Gaussian filtering. The image was mirrored along its boundaries before contrast enhancement to avoid artifacts, and the region outside its boundaries was blacked out after enhancement. 
\end{enumerate}

The geometric steps (cropping and resizing) were replicated on the labels for training. Figure \ref{fig:prep} shows outputs of our pre-processing algorithm. Models were trained on the concatenation of the cropped CFI and its contrast-enhanced version (6 input channels).

\subsubsection*{Vessel, A/V and Disc Segmentation Models}

Vessel and A/V segmentation models were trained separately given the different development sets. Data was augmented via random appearance transforms: de-focus, hue-saturation-value, Gamma and Gaussian noise; followed by geometric transforms: flipping (horizontal and vertical), random scale (1.0-1.15 times original scale) random rotation (up to 10 degrees), random elastic transforms and finally random cropping of a $512x512px$ patch. This was done to maintain enough resolution for the segmentation of the thinner vessels. Correspondingly, inference is done on $1024x1024px$ images via a sliding window with 50\% overlap. A Gaussian kernel was used for merging window segmentations. Test-time augmentation was also applied: the outputs of horizontal, vertical, and horizontal-vertical flips were averaged together before binarizing the final output. These procedures closely follow or are inspired by the process followed in NN-Unet \cite{nnunet}.

Vessel annotations were encoded as 0) background and 1) vessel. Artery-vein annotations were encoded as 0) background, 1) arteries, 2) veins, and 3) crossings (both artery and vein). We used this approach for A/V segmentation with the goal of boosting the relative importance of the crossings in the loss function. The loss function was the sum of Dice and Cross-entropy with equal weights. The optic disc segmentation model was similarly trained on binary masks encoded as 0) background and 1) disc mask; with the same loss function. Figure \ref{fig:batch_vessels} shows sample training batches for the vessel and artery-vein segmentation models.

The models were trained making use of a custom implementation of the U-Net with 6 input channels (for original RGB image and contrast-enhanced version), 8 down-sampling stages, and deep supervision (loss calculated on the down-sampled stages too). The input resolution was $512x512px$.

\subsubsection*{Fovea Localization Model}

The fovea localization model was trained separately to localize keypoints on the CFIs. A heatmap regression approach \citep{heatmap-regression} with the same U-Net architecture used for artery-vein segmentation as backbone. Heatmap regression consists in training the model to regress a probability image or heatmap of the keypoint location. The pixel location of the maximum on the output heatmap is then taken as the output keypoint location. Train heatmaps may be generated using different probability density functions centered at the ground truth keypoint location. We generated target heatmaps ($512x512px$) using a Gaussian with constant $\sigma$ centered at the keypoint. The value of $\sigma = 50px$ was selected via experiments on a holdout set for the task of localizing the disc center on CFIs. Mean squared error (MSE) was used as loss function to train the model to approximate the continuous target heatmaps. The location (2D-index) of the maximum over the heatmaps output by the model was taken as the keypoint location.

Input CFIs were augmented via random appearance transforms: de-focus, hue-saturation-value, Gamma and Gaussian noise; followed by geometric transforms: horizontal flipping, random scale (1.0-1.05 times original scale) random rotation (up to 5 degrees) and finally resizing down to model input size. Test-time augmentation was correspondingly applied using the vertical flip of the input. Figure \ref{fig:batches_disc_fovea} shows sample training batches for the disc segmentation and fovea localization models.

\subsubsection*{Quality Classification Model}

The image quality classification model was trained using the same augmentations as for fovea and disc models. The models were developed by fine-tuning Resnet101 (Imagenet weights) with $224px$ input resolution. Here we used only the original RGB CFI (3 channels; no contrast-enhanced version). All the model weights were fine-tuned (no frozen layers). The cross-entropy loss was used for classification and Mean Squared Error (MSE) for regression. Test-time augmentation was applied by averaging the outputs of the original and horizontally-flipped inputs.

\subsubsection*{Training Setup}

For all models, we made use of the \textit{albumentations} library for data augmentations. Pytorch and \textit{pytorch-lightning} were used to develop our model training pipeline.

Training and evaluation were done via a 5-fold cross-validation loop. Group-fold cross-validation was used to ensure that images of the same eye were always together in either training or test set. The batch size was set to 16 for all models. The \textit{Adam} optimizer was used with $lr=0.001$ for segmentation. Models were trained for a fixed number of epochs across folds: 100 for vessel segmentation; 200 for A/V segmentation; 100 for disc segmentation and 35 for fovea localization. This was decided approximately based on experiments with a partition of 20\% of the training set. We observed flat performance progression on these sets before and after the chosen epoch number and did not consider it necessary to do further optimization. Some architectural choices such as the use of deep supervision and increased UNet depth were validated using the same partition.

Per-fold models were trained on a cluster using a single NVidia graphics card per model. Segmentation models were trained in mixed precision. 

\subsection*{Evaluation}

We evaluated our models via 5-fold cross-validation, with the Dice score as the main evaluation metric. We report the mean scores (separated per source dataset) over the five validation sets. This is to provide a score of generalization performance on all datasets \ref{tab:datasets}. For all models, we input pre-processed images in $1024x1024px$ resolution. Test-time augmentation was enabled. Note that this evaluation setup means that the scores of our models could be slightly under-represented when compared to the full ensemble of our models.

For the A/V model artery and vein masks were recovered from the model outputs. For the VascX model, this meant merging the \textit{Crossings} output mask with both \textit{Artery} and \textit{Vein} masks.






    


\section*{Results}

\subsection*{Comparison with Publicly Available Models}
\label{seg:main_comp}

We compared the performance of our models with that of publicly available ones. Note that our goal was not to compare deep learning architectures but instead to compare the performance of entire systems (including architecture, inference code, and public model weights) across different datasets (Table \ref{tab:datasets}). Therefore architectures published without publicly available weights have been excluded from this analysis.

We compared the three segmentation models (vessels, artery-vein, disc) to the models in Automorph \citep{automorph}, the most comparable pipeline that integrates vessel, A/V, and disc segmentation. For each task, we compared against the corresponding model ensemble (7 models) in the Automorph pipeline. Additionally, we compared vessel and A/V models to Little-WNet (LWNet) \citep{lwnet}, a dedicated vessel and A/V segmentation model with publicly available weights. The LWNET vessel segmentation weights were trained on DRIVE. For A/V segmentation we tested both DRIVE and HRF-trained weights. We found DRIVE weights to perform better across all datasets and therefore report only these scores.

\begin{table}[!p]
    \centering

    \begin{subtable}[b]{\textwidth}
    \centering
    \begin{tabular}{l*{7}{c}}
        \toprule
        Model & ChaseDB & DRHAGIS & HRF & RETA & FIVES & Leuven-Haifa & Rotterdam \\
        \midrule
        Automorph & NA & .741 & NA & NA & .824 & .812 & .810 \\
        LWNet & .732 & .672 & .573 & NA & .756 & .713 & .745 \\
        VascX & \textbf{.780} & \textbf{.769} & \textbf{.786} & \textbf{.890} & \textbf{.831} & \textbf{.829} & \textbf{.852} \\

        \bottomrule
    \end{tabular}
    \caption{Vessel segmentation (Dice score). }
    \end{subtable}
    \vspace{0.5cm}

    \begin{subtable}[b]{\textwidth}
    \centering
    \begin{tabular}{l*{10}{c}}
        \toprule
        \multirow{2}{*}{Model} & \multicolumn{2}{c}{HRF-AV} & \multicolumn{2}{c}{RITE}  & \multicolumn{2}{c}{Les-AV} & \multicolumn{2}{c}{Leuven-Haifa} & \multicolumn{2}{c}{Rotterdam} \\
        \cmidrule{2-11}
         & A & V & A & V & A & V & A & V & A & V \\
        \midrule
        Automorph \citep{automorph} & NA & NA & NA & NA & NA & NA & .707 & .773 & .712 & .752 \\
        LWNet \citep{lwnet} & .506 & .574 & NA & NA & .618 & .670 & .569 & .649 & .635 & .696 \\
        VascX & \textbf{.746} & \textbf{.792} & \textbf{.711} & \textbf{.760} & \textbf{.800} & \textbf{.835} & \textbf{.786} & \textbf{.826} & \textbf{.817} & \textbf{.849} \\
        \bottomrule
    \end{tabular}
    \caption{Artery-vein segmentation (Dice score). A: arteries mask score, V: veins mask score}
    \end{subtable}
    \vspace{0.5cm}

    \begin{subtable}[b]{\textwidth}
    
    \centering
    \begin{tabular}{l*{8}{c}}
        \toprule
        Model &  ORIGA & PAPILA & IDRiD & ADAM & PALM & REFUGE2 & Rotterdam \\
        \midrule
        Automorph & .876 & .618 & .931 & .958 & .846 & NA & .220 \\
        VascX & \textbf{.958} & \textbf{.961} & \textbf{.958} & \textbf{.964} & \textbf{.921} & \textbf{.956} & \textbf{.886} \\ 
        \bottomrule
    \end{tabular}
    
    \caption{Disc segmentation (Dice score)}
    \label{tab:comparison_disc}
    \end{subtable}
    \vspace{0.5cm}

    \begin{subtable}[b]{\textwidth}
    \centering
    \begin{tabular}{l*{6}{c}}
        \toprule
        Model & IDRID & ADAM & PALM & REFUGE2 & Rotterdam \\
        \midrule
        VascX & 14.68 & 11.15 & 26.22 & 10.68 & 25.14 \\
        \bottomrule
    \end{tabular}
    \caption{Fovea localization (L2/Euclidean distance to ground truth in pixels)}
    \end{subtable}
    
    \caption{Performance comparison of VascX segmentation models. \textit{NA} is used for a model - dataset pair when the model was trained on the dataset and therefore it was not possible to benchmark this model - dataset pair on out-of-distribution data.}
    \label{tab:comparison}
\end{table}

Table \ref{tab:comparison} shows the results of our comparison using Dice score (segmentation) and Euclidean distance (fovea localization) as evaluation metrics, which reveal consistently higher performance from VascX across all datasets. Note that not all model - dataset pairs were evaluated due to the model being trained on the corresponding dataset entirely.

For A/V segmentation VascX achieved higher Dice scores across all datasets, with slightly higher performance for veins than for arteries, as has been observed in previous work \citep{lunet}. Importantly, we observed large improvements in the Rotterdam and Leuven-Haifa datasets, the only ones evaluated across the three models (Figure \ref{fig:ds_stats}). The differences was $\sim0.1$ for the Rotterdam dataset, the most diverse of the datasets; suggesting increased model robustness.

Regarding disc segmentation, the largest difference was in the Rotterdam dataset, where Automorph achieved a score of only 0.220. Many of these failure cases were due to Automorph's bounds detection algorithm failing on Rotterdam CFIs. The disc detection was however also prone to fail for discs located close to or crossing the bounds of the CFI; and for images for which the disc border had low contrast. 

\begin{figure}[!ht]
    \centering
    \includegraphics[width=0.9\textwidth]{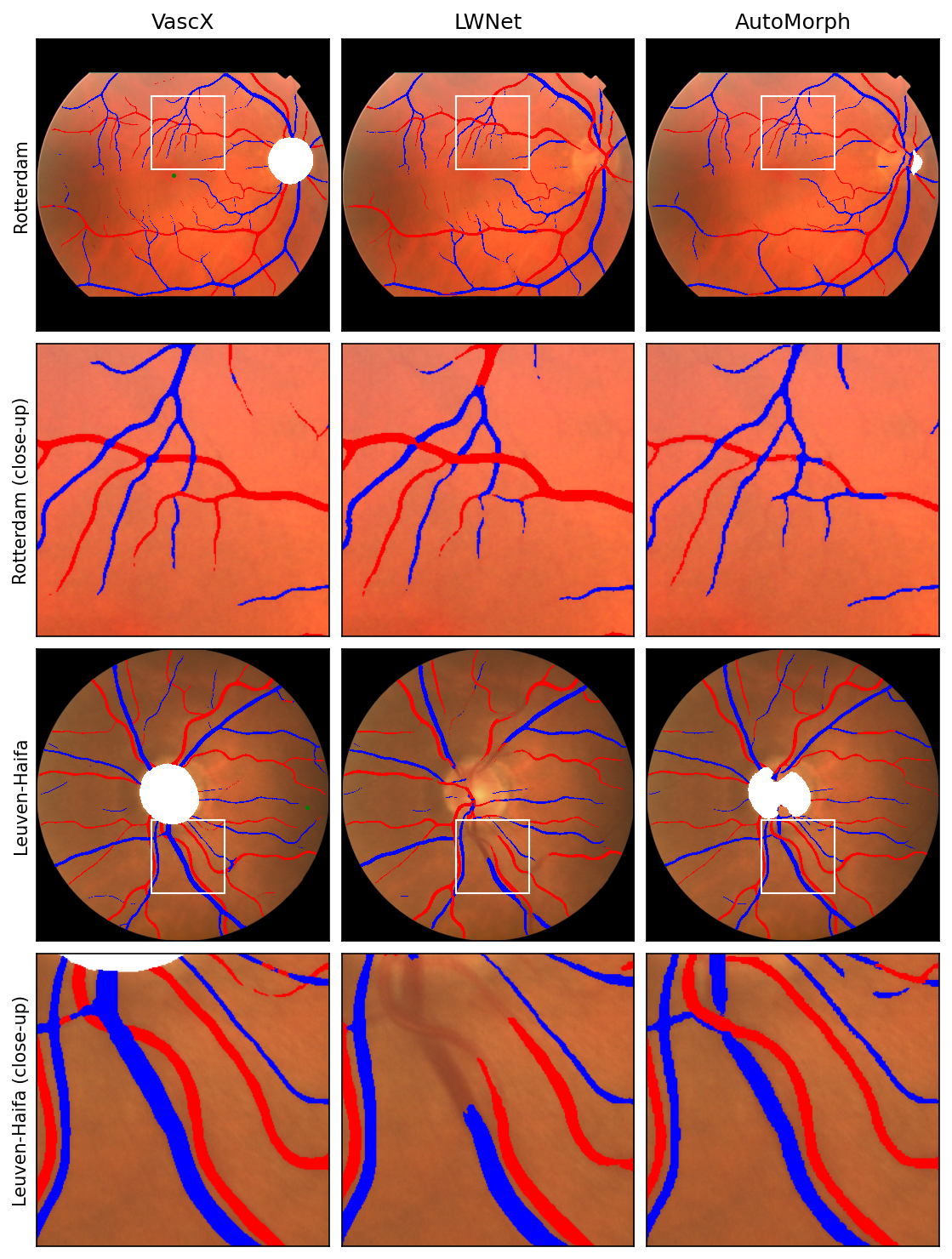}
    \caption{Example model outputs from the benchmarked systems showing A/V segmentation and disc and fovea outputs were applicable. We observed more consistency in A/V and disc segmentation from VascX. The CFIs shown correspond to the images with the median Dice score for A/V segmentation with VascX; for each dataset.}
    \label{fig:medians}
\end{figure}

Qualitatively we observed a marked improvement in segmentation performance from VascX models. Figure \ref{fig:medians} shows example model outputs from Rotterdam and Leuven-Haifa inputs (the two independent datasets). We observed more consistent disc detection from VascX, especially in cases where the disc was not clearly differentiated. A/V segmentation, while not flawless, produced visibly less continuity and mis-classification mistakes than both LWNet and Automorph. Importantly, many of the vessel crossings were correctly resolved on both artery and vein masks.

\subsection*{Effect of Image Quality}

\begin{figure}[ht]
    \centering
    \begin{subfigure}[b]{0.55\textwidth}
        \centering
        \includegraphics[height=12cm]{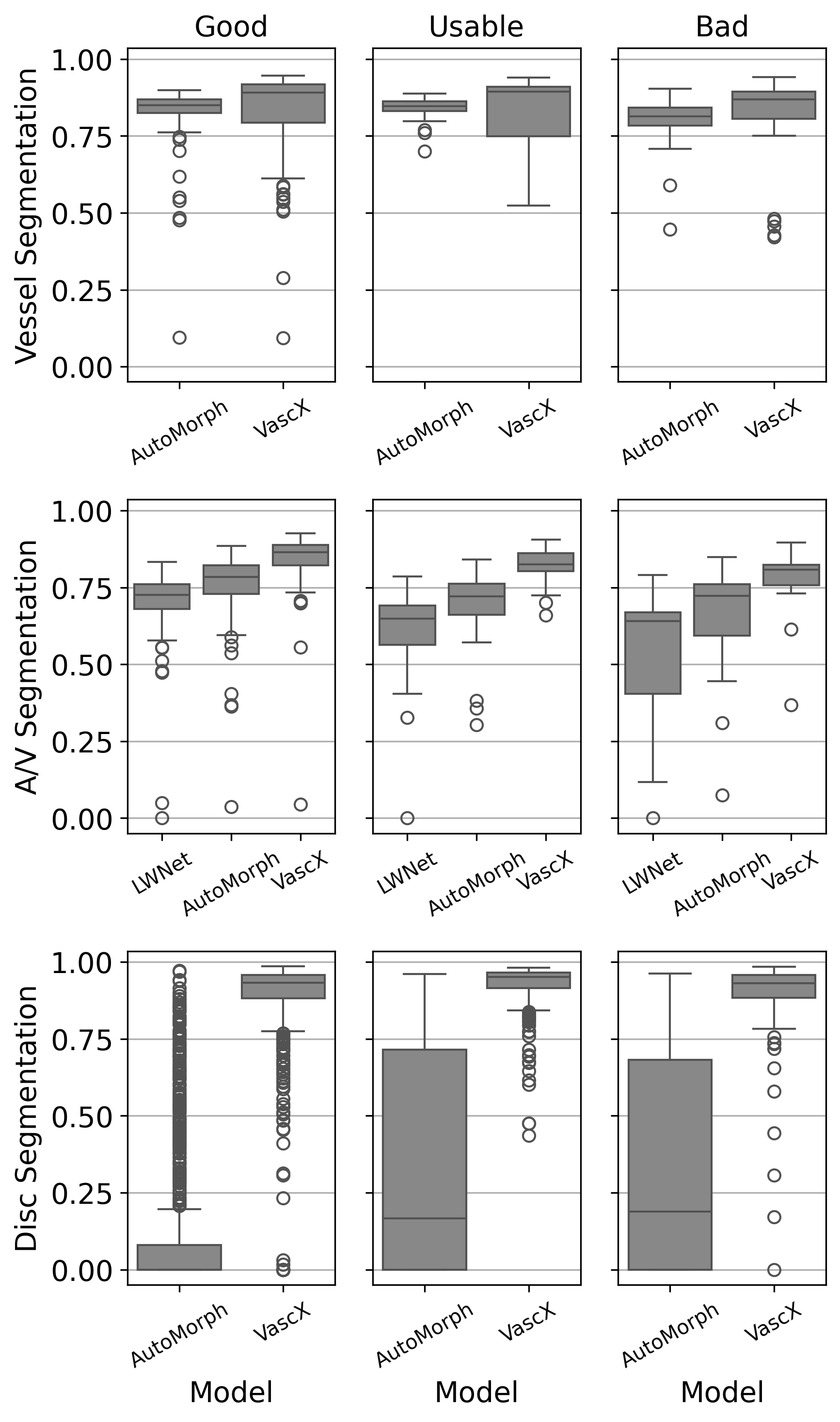}
        \caption{CFI quality.}
        \label{fig:boxplots_quality}
    \end{subfigure}
    \hfill
    \begin{subfigure}[b]{0.35\textwidth}
        \centering
        \includegraphics[height=12cm]{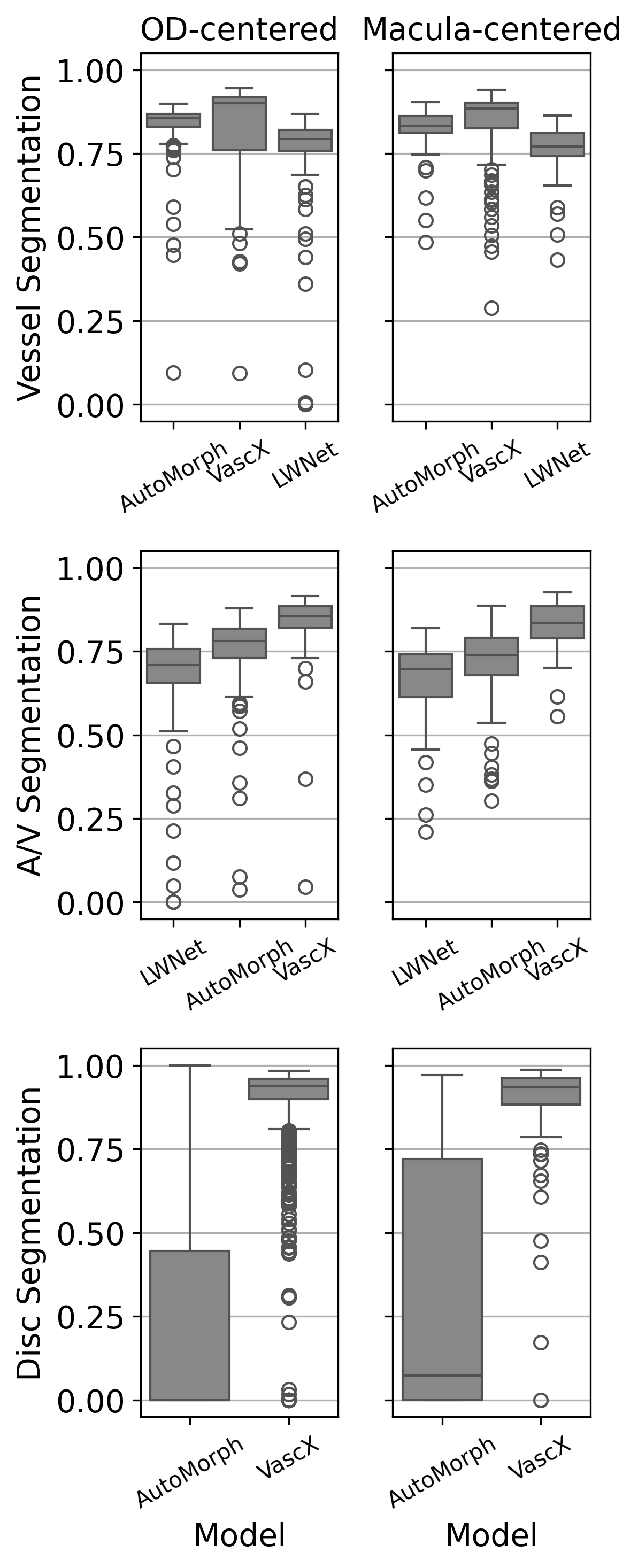}
        \caption{Anatomic region.}
        \label{fig:boxplots_ar}
    \end{subfigure}
    \caption{Box plots displaying the distribution of model performance (Dice score) over image quality and anatomic region (AR) bins/slices of the Rotterdam dataset. For vessel segmentation, the performance of the models is stable across quality and anatomic region bins. For artery-vein segmentation, the biggest improvement is in \textit{usable} images. For disc segmentation, the model achieves remarkably stable performance across image quality and AR; while the images pose a challenge for Automorph.}
    \label{fig:boxplots}
\end{figure}

To more deeply characterize the performance of the model, we made use of the CFI quality estimation model to classify Rotterdam images into \textit{Good}, \textit{Usable} and \textit{Bad / Unusable} categories. We limited this analysis to the Rotterdam dataset set due to the other dataset being smaller and/or highly uniform in terms of image quality and capture conditions. Figure \ref{fig:boxplots_quality} displays the distribution of Dice scores over image quality bins, as assessed by the model. Figure \ref{fig:boxplots_ar} shows the Dice score distributions for optic disc and macula-centered images. Once again vessel segmentation models achieved high consistency even for CFIs classified as \textit{bad} quality.

\subsection*{Feature Accuracy}

The primary goal of our models is to serve as the backbone for explainable vascular feature/biomarker extraction systems. Therefore we quantified the quality of such features. We compared features extracted from model segmentations (VascX, Automorph (AM) and LittleWNet (LWN)) with reference features extracted from the annotations in our development set, which we considered ground truth.
. We quantified the quality of model features by measuring mean absolute error (MAE) and correlations with the ground truth features. We used the Rotterdam set as main evaluation set for this comparison due to its diversity and size. Due to our 5-fold cross-validation setup for training, for VascX we made use of segmentations generated within each fold to have independence of the test set. Therefore the performance of our models might be under-estimated when compared to the full ensemble.

We made use of a vascular feature extraction to extract a set of features including tortuosity and central retinal equivalents (CRE) from segmentation masks. See Supplementary Section \ref{app:feature_details} for feature implementation details. The features are calculated using artery-vein segmentation masks, and some (CREs, temporal angles, vascular density) use disc segmentation masks and fovea locations to define regions. Because LWNet does not include optic disc segmentation weights and Automorph struggled with the disc segmentation of Rotterdam images (Table \ref{tab:comparison_disc}), we used the same ground truth disc masks in both pipelines. The fovea location was also the same across models. Therefore, this comparison evaluates the effect of different models for artery-vein segmentation only. 

Table \ref{tab:features} shows the results of the comparison, where we measured reduced mean absolute error and higher correlations when using VascX segmentations, across most features. Of 24 vascular features evaluated, 16 showed a significant improvement in MAE when compared to Automorph, and 22 when compared to LWNet. VascX also had the highest correlations with ground truth features in all but one case.

We report results of the same comparison on the Leuven-Haifa dataset (240 images with $30\degree$ FOV) in Table \ref{tab:features_leuven}. These results show similar behavior. Although less features improved over Automorph in terms of MAE, all correlations favoured VascX.

\begin{table}[!ht]
\centering
\begin{tabular}{rr|rrr|rrr}
\toprule
 & \multirow{2}{*}{Grader $\mu$} & \multicolumn{3}{c}{MAE} & \multicolumn{3}{c}{Pearson \textit{r}} \\
 & & VascX & AM & LWN & VascX & AM & LWN \\
\midrule
Temporal Angle - A & 122.36 & \textbf{5.56} & 6.39* & 8.63$\dagger$ & \textbf{0.69} & 0.66 & 0.57 \\
Temporal Angle - V & 126.35 & \textbf{5.05} & 5.91 & 8.29$\dagger$ & \textbf{0.77} & 0.73 & 0.65 \\
CRE - A & 10.53 & \textbf{0.93} & 1.49$\dagger$ & 3.10$\dagger$ & \textbf{0.79} & 0.57 & 0.49 \\
CRE - V & 15.83 & \textbf{1.11} & 1.94$\dagger$ & 3.79$\dagger$ & \textbf{0.82} & 0.70 & 0.51 \\
Vasc. Density - A & 4.25 & \textbf{0.33} & 0.85$\dagger$ & 0.76$\dagger$ & \textbf{0.89} & 0.76 & 0.61 \\
Vasc. Density - V & 4.90 & \textbf{0.35} & 0.87$\dagger$ & 1.12$\dagger$ & \textbf{0.88} & 0.64 & 0.60 \\
Vessel Caliber [med] - A & 4.23 & \textbf{0.60} & 0.64 & 1.36$\dagger$ & \textbf{0.63} & 0.48 & 0.37 \\
Vessel Caliber [std] - A & 2.26 & \textbf{0.27} & 0.39$\dagger$ & 0.90$\dagger$ & \textbf{0.70} & 0.59 & 0.49 \\
Vessel Caliber [med] - V & 4.35 & 0.83 & \textbf{.72}* & 1.38$\dagger$ & 0.49 & \textbf{0.54} & 0.38 \\
Vessel Caliber [std]- V & 3.15 & \textbf{0.26} & 0.47$\dagger$ & 0.79$\dagger$ & \textbf{0.84} & 0.69 & 0.64 \\
Tortuosity [med] - A & 1.09 & \textbf{0.0058} & 0.006 & 0.0059 & \textbf{0.81} & 0.73 & 0.76 \\
Tortuosity [med] - V & 1.09 & \textbf{0.0048} & 0.0053 & 0.0059* & \textbf{0.71} & 0.57 & 0.52 \\
Curvature [med] - A & 9.08 & \textbf{1.07} & 1.27* & 1.34* & \textbf{0.93} & 0.90 & 0.86 \\
Curvature [med] - V & 10.25 & \textbf{1.11} & 1.40$\dagger$ & 1.44$\dagger$ & \textbf{0.84} & 0.84 & 0.83 \\
Inflection count [med] - A & 1.46 & \textbf{0.48} & 0.61* & 0.53 & \textbf{0.39} & 0.26 & 0.30 \\
Inflection count [med] - V & 1.57 & \textbf{0.45} & 0.50 & 0.60* & \textbf{0.43} & 0.32 & 0.26 \\
Tortuosity [LW] - A & 1.12 & \textbf{0.017} & 0.019* & 0.038$\dagger$ & \textbf{0.52} & 0.47 & 0.11 \\
Tortuosity [LW] - V & 1.11 & \textbf{0.012} & 0.014* & 0.033$\dagger$ & \textbf{0.75} & 0.62 & 0.12 \\
Bif. Angles [mean] - A & 82.99 & \textbf{5.79} & 6.70* & 9.51$\dagger$ & \textbf{0.48} & 0.41 & 0.27 \\
Bif. Angles [med] - A & 81.83 & \textbf{5.19} & 6.15 & 8.97$\dagger$ & \textbf{0.57} & 0.45 & 0.22 \\
Bif. Angles [mean] - V & 84.16 & \textbf{4.47} & 5.53* & 7.71$\dagger$ & \textbf{0.75} & 0.59 & 0.54 \\
Bif. Angles [med] - V & 83.20 & \textbf{3.75} & 5.02* & 5.96$\dagger$ & \textbf{0.74} & 0.52 & 0.56 \\
Num. Bifurcations - A & 27.32 & \textbf{10.34} & 10.97 & 12.59$\dagger$ & \textbf{0.81} & 0.77 & 0.69 \\
Num. Bifurcations - V & 31.61 & \textbf{10.33} & 12.06$\dagger$ & 13.48$\dagger$ & \textbf{0.89} & 0.76 & 0.76 \\
    \bottomrule
    \end{tabular}
    \caption{Feature quality comparison on the Rotterdam sets (215 CFIs). Results of comparing MAE of features extracted using different models for artery-vein segmentation: VascX, Automorph (AM) and LittleWNet (LWN). Features extracted from ground truth segmentations were used as the reference for MAE. The feature implementation was the same across models. For vessel or bifurcation features, the aggregation function (applied on features from the whole image) is indicated in brackes: [med]: median, [std]: standard deviation, [mean]: mean value, [LW]: length-weighted; vessels were weighted by their length. Grader $\mu$ is the mean of feature values; mean MAE is the mean absolute error between ground truth features and features extracted from model outputs. Pearson coefficients are correlations between ground truth and model outputs. *: Significant difference in MAE ($p < 0.05$) between the model and VascX. $\dagger$: $p < 0.001$}
    \label{tab:features}
\end{table}

\section*{Discussion}
We presented new model ensembles for CFI vasculature analysis and benchmarked against publicly available systems. With the objective of developing models that operate robustly across devices and achieve performance improvements on the different imaging conditions found on real cohort studies and clinical datasets, we focused on the diversity of our development set. We augmented more than 15 public datasets (each of which usually consist of carefully selected CFIs from one or a few devices) with CFIs collected in The Netherlands throughout more than three decades from 11 different imaging devices and with diverse pathology. Together, this resulted in a comprehensive set of images including a wider range of patient characteristics (origin, age, pathologies) and capture characteristics (device, anatomic region, field of view). For model training, we made use of well-known architectures, a new more robust pre-processing algorithm, and data augmentation to achieve solid performance across tasks.

Our main evaluation using Dice scores revealed consistently better performance from VascX when compared to existing publicly available models - Automorph and LWNET - across both public datasets and the Rotterdam sets. We observed an important improvement in A/V segmentation performance, crucial in vascular analysis pipelines. It stands out that the largest difference between VascX and the rest is in \textit{usable} or intermediate quality images \ref{fig:boxplots}, common in large cohort studies and clinical data. A large improvement was observed in disc segmentation performance when compared to Automorph, which struggled with Rotterdam images likely due to its more limited training set \citep{automorph}. Its performance was low for both disc and macula-centered images and across quality levels (\ref{fig:boxplots}). A large difference was also observed in PAPILA \ref{tab:comparison}, which has a 30 degree field of view not present in Automorph's training set. The differences in vessel segmentation performance were smaller between models, suggesting that performance in this task may already be saturated.

Our fovea localization model had mean L2 errors between 10.68 (REFUGE2) and 26.22 (PALM) on images of $1024\times1024 px$. This model is enough to provide an approximate localization of the fovea center, useful for the automatic placement of retinal regions (eg. ETDRS grid) for localized measurements.

Importantly, our results showed that better segmentation performance translates into more accurate vascular features, in line with the results of \citet{lunet}. Features from VascX model segmentations show lower absolute error and higher correlation with features extracted directly from grader segmentations; when compared to features extracted from Automorph segmentations.

Further work is required in several areas. There is room for improvement in artery-vein segmentation, especially regarding the connectivity of the vessel trees. Despite the improvements, our model still produces A/V mis-classifications and gaps. Enforcing the expected tree structure in the segmentation is a notorious open challenge in deep learning. Incorporating specific loss functions \citep{cldice} and post-processing steps \citep{greco} designed for connectivity may help address this issue.

The further development of retinal analysis pipelines and the development and evaluation of vascular features with clinical significance also holds promise. The feature computation stage involves significant nuance. Some features such as tortuosity, for example, may be particularly sensitive to discontinuities in the segmentation. The results in Table \ref{tab:features} support that some features are more robust than others to potential mistakes in the segmentations. Deeper evaluations of the robustness and predictive power of vascular features are in order. The ability to capture image quality for quality control and for informing feature computation may be crucial here. Classification of image quality using a one-dimensional image-level label, while useful, is not granular enough for some applications. Artifacts may affect only part of the image. Defocus or poor overall quality may result in a CFI usable for analysis of the main vascular arches but not any smaller vessels. A decomposition of image quality into (localized) factors may be a necessary next step in developing more robust pipelines.



In summary, our results show that our models outperform those in previous systems in every dataset and condition we evaluated, likely due to the increased size and variety of our training set. We expect that VascX models will serve as the backbone for more robust CFI vascular analysis pipelines (and other CFI-based pipelines). To serve these goals, we have made available our entire inference pipeline, including model weights and inference code with a new CFI pre-processing algorithm. With this, we aim to catalyze research towards understanding how analysis of the retinal vasculature can help us prevent, detect and monitor diseases, not only of the eye but of the rest of the body.



\section*{Acknowledgments}

The authors acknowledge the staff of the Eyened Reading Center for their different roles in data collection and algorithm development. Special acknowledgment to the Sinergia consortium for conceiving and supporting this project. The authors are grateful to the study participants and the staff from the Rotterdam Study, AMD-Life; the Dutch Myopia Study and to all contributors to the (publicly available) datasets that we used to train our models. This work was funded by the Swiss National Science Foundation grant no. CRSII5_209510.

\bibliography{sample}

\newpage

\appendix

\section{Dataset details}

\begin{table}[h!]
    \centering
    \captionsetup{justification=centering}
    
    \caption{Complementary details of the public dataset used in model development. *: 379 of 1200 samples were discarded due to containing no disc segmentation for partially visible discs. $\dagger$: 21 of 1200 CFIs were discarded due to having no associated annotation.}
    \begin{tabular}{>{\arraybackslash}p{3cm}>{\arraybackslash}p{1cm}>{\arraybackslash}p{4.5cm}>{\arraybackslash}p{4.5cm}}
        \toprule
        \textbf{Dataset} & \textbf{N} & \textbf{Resolution}  & \textbf{Devices} \\
        \vspace{1pt} \\
        \multicolumn{4}{c}{\textbf{Vessel Segmentation}} \\
        \hline
        \noalign{\vskip 5pt} 

        Chase DB \citep{CHASEDB1} & 28 & 35\degree, $1280\times960$ & Nidek NM-200-D \\
        DRHAGIS \citep{drhagis} & 40 & 45\degree, $4752\times3168 - 2816\times1880$ & Topcon TRC-NW6s, Topcon TRC-NW8, Canon CR DGi \\
        HRF \citep{hrf} & 45 & 45\degree, $3504\times2336$ & NR \\
        RETA \citep{reta} & 54 & 50\degree, $4288\times2848$ & Kowa VX-10 alpha \\
        FIVES \citep{fives} & 800 & 50\degree, $2048\times2048$ & Topcon TRC-NW8 \\
        Leuven-Haifa \citep{leuven} & 240 & $1444\times1444$ & Zeiss Visucam 500 \\
        
        Rotterdam (ours) & 352 & \multicolumn{2}{c}{see Fig. 1} \\
        \noalign{\vskip 5pt}
        \textbf{Total} & \textbf{1559} & & \\
        
        \vspace{0.4pt} \\
        \multicolumn{4}{c}{\textbf{Artery Vein Segmentation}} \\
        \hline
        \noalign{\vskip 5pt}
        
        RITE \citep{avdrive} & 40 & 45\degree, $768\times584$ & Canon CR5 3CCD \\
        HRF-AV \citep{hrfav} & 45 & 45\degree, $3504\times2336$ & NR \\
        Les-AV \citep{lesav} & 22 & 30\degree, $1444\times1620$ & NR \\
        Leuven-Haifa \citep{leuven} & 240 & 30\degree, $1444\times1444$ & Zeiss Visucam 500 \\
        Rotterdam (ours) & 215 & \multicolumn{2}{c}{see Fig. 1} \\
        \noalign{\vskip 5pt}
        \textbf{Total} & \textbf{562} & & \\

        \vspace{0.4pt} \\
        \multicolumn{4}{c}{\textbf{Disc Segmentation}} \\
        \hline
        \noalign{\vskip 5pt}

        ORIGA \citep{origa} & 650  & NR & NR \\
        PAPILA \citep{papila} & 488 & 45 \degree, $2576\times1934$ & Topcon TRC-NW400 \\
        IDRiD \citep{idrid} & 81 & 50\degree, $4288\times2848$ & Kowa VX-10 alpha \\
        ADAM \citep{adam} & 821* & $2124\times2056$, $1444 \times 1444$ & Zeiss Visucam 500, Canon CR-2  \\ 
        PALM \citep{palm} & 1179$\dagger$ & $2124\times2056$, $1444 \times 1444$ & Zeiss Visucam 500, Canon CR-2 \\ 
        REFUGE2 \citep{refuge} & 2000 & $1634 \times 1634$ - $2124 \times 2056$ &  Zeiss Visucam 500, Canon CR-2, Topcon TRC-NW400, Kowa \\
        Rotterdam (ours) & 1225 & \multicolumn{2}{c}{see Fig. 1} \\
        \noalign{\vskip 5pt}
        \textbf{Total} & \textbf{7464} & & \\

        \vspace{0.4pt} \\
        \multicolumn{4}{c}{\textbf{Fovea Localization}} \\
        \hline
        \noalign{\vskip 5pt}
        
        IDRiD \citep{idrid} & 516 & 50\degree, $4288\times2848$ & Kowa VX-10 alpha \\
        ADAM \citep{adam} & 1200 & $2124\times2056$, $1444 \times 1444$ & Zeiss Visucam 500, Canon CR-2  \\
        PALM \citep{palm} & 1200 & $2124\times2056$, $1444 \times 1444$ & Zeiss Visucam 500, Canon CR-2 \\
        REFUGE2 \citep{refuge} & 2000 & $1634 \times 1634$ - $2124 \times 2056$ &  Zeiss Visucam 500, Canon CR-2, Topcon TRC-NW400, Kowa \\
        
        Rotterdam (ours) & 10908 &  \multicolumn{2}{c}{see Fig. 1}  \\
        \noalign{\vskip 5pt}
        \textbf{Total} & \textbf{15824} & & \\

        \vspace{0.4pt} \\
        \multicolumn{4}{c}{\textbf{CFI Quality Estimation}} \\
        \hline
        \noalign{\vskip 5pt}

        EyeQ \citep{eyeq} & 28,792 & &  \\
        \hline
    \end{tabular}
    \label{tab:datasets_appendix}
\end{table}

\newpage

\section*{Sample training batches}

\begin{figure}[ht!]
    \centering
    \begin{subfigure}[b]{1.0\textwidth}
        \centering
        \includegraphics[width=\textwidth]{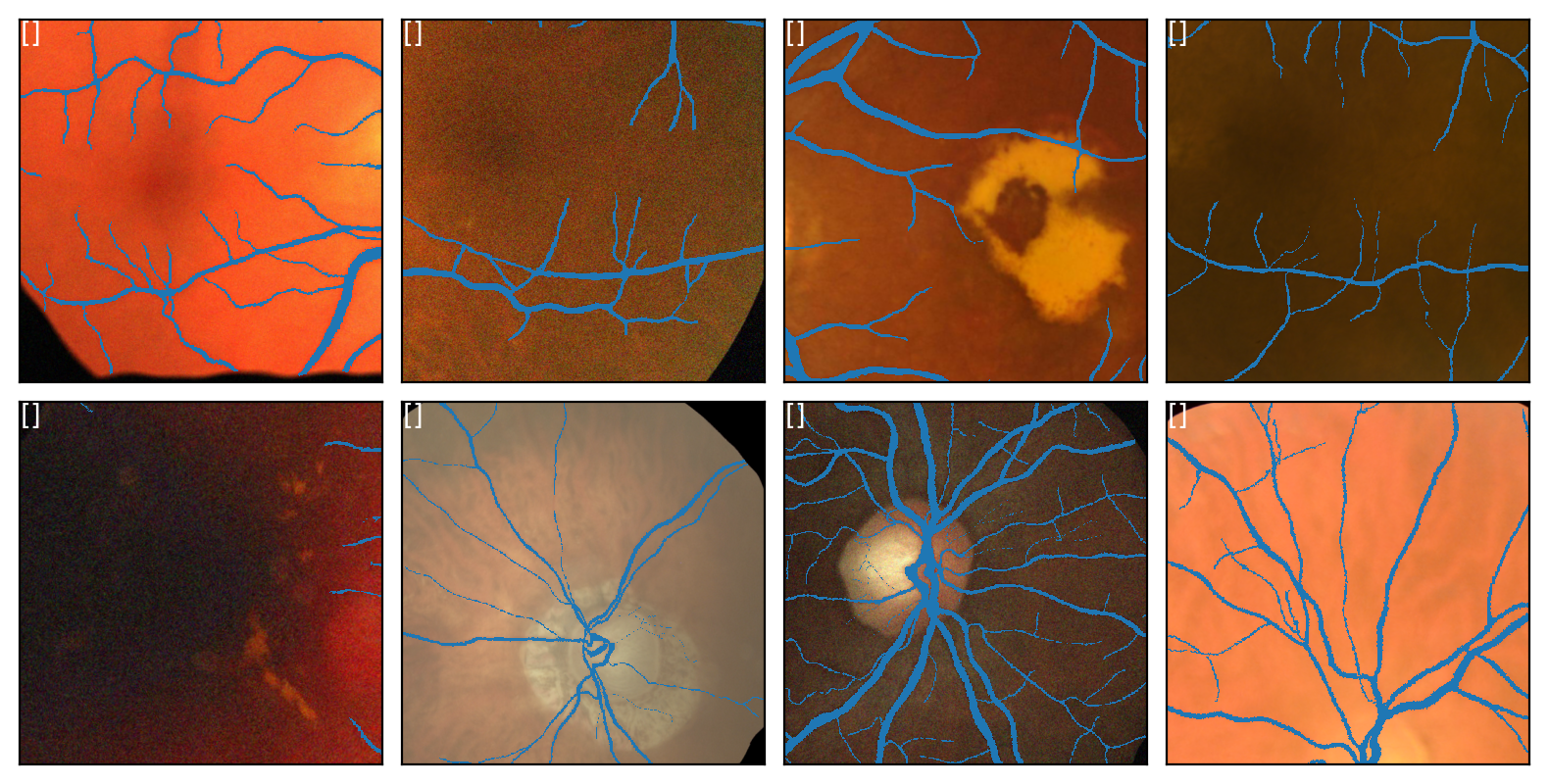}
        \caption{Vessel segmentation.}
        \label{fig:batch_vessels}
    \end{subfigure}
    \hfill
    \begin{subfigure}[b]{1.0\textwidth}
        \centering
        \includegraphics[width=\textwidth]{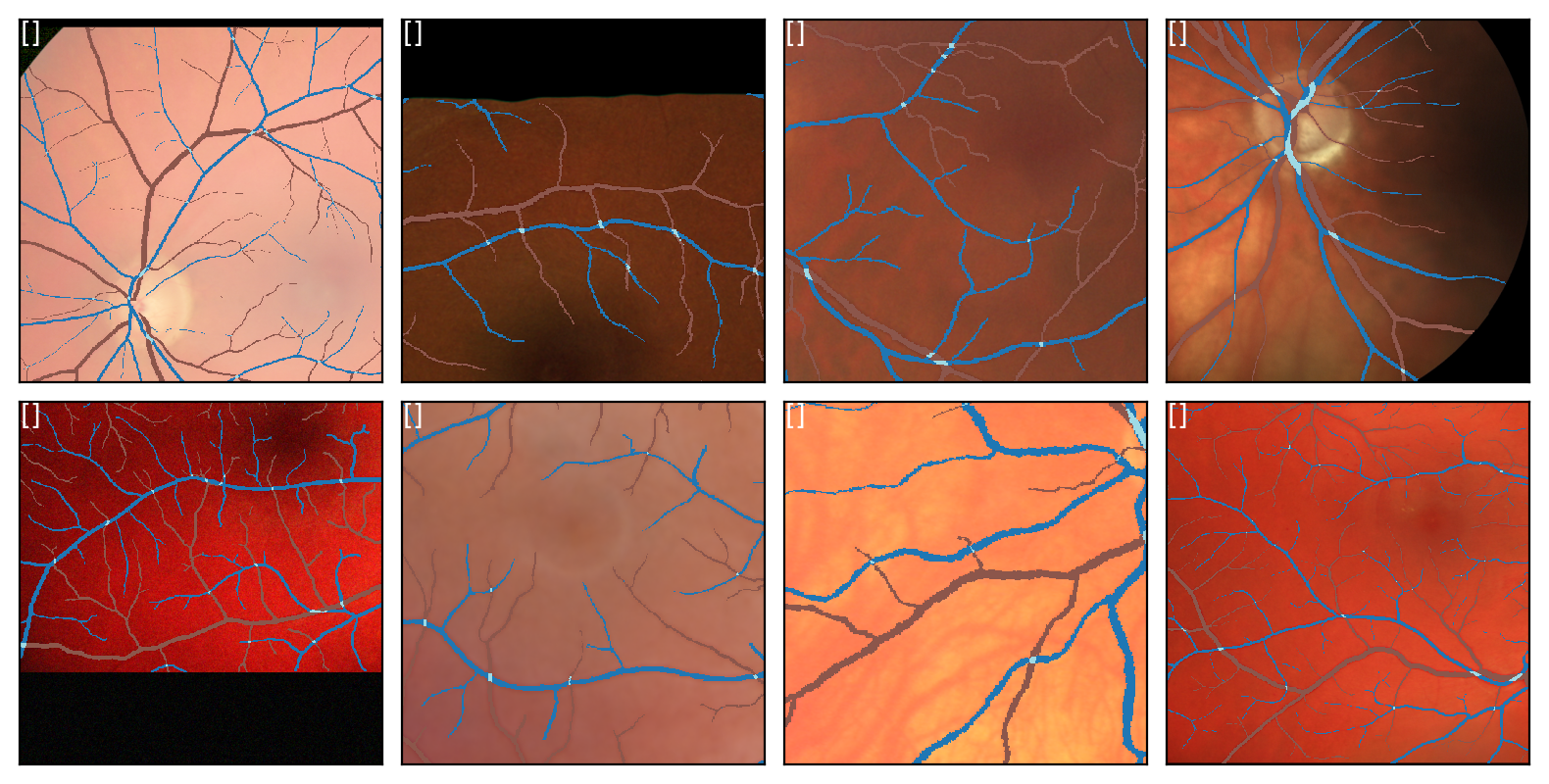}
        \caption{Artery-vein segmentation, where arteries, veins and crossings are separate (exclusive) classes.}
        \label{fig:batch_av}
    \end{subfigure}
    \caption{Sample training batches of input images to vessel and artery-vein segmentation models after data augmentation, including the CFI and overlaid masks. The contrast enhanced image is not shown.}
    \label{fig:batches_av_vessels}
\end{figure}

\begin{figure}[ht!]
    \centering
    \begin{subfigure}[b]{1.0\textwidth}
        \centering
        \includegraphics[width=\textwidth]{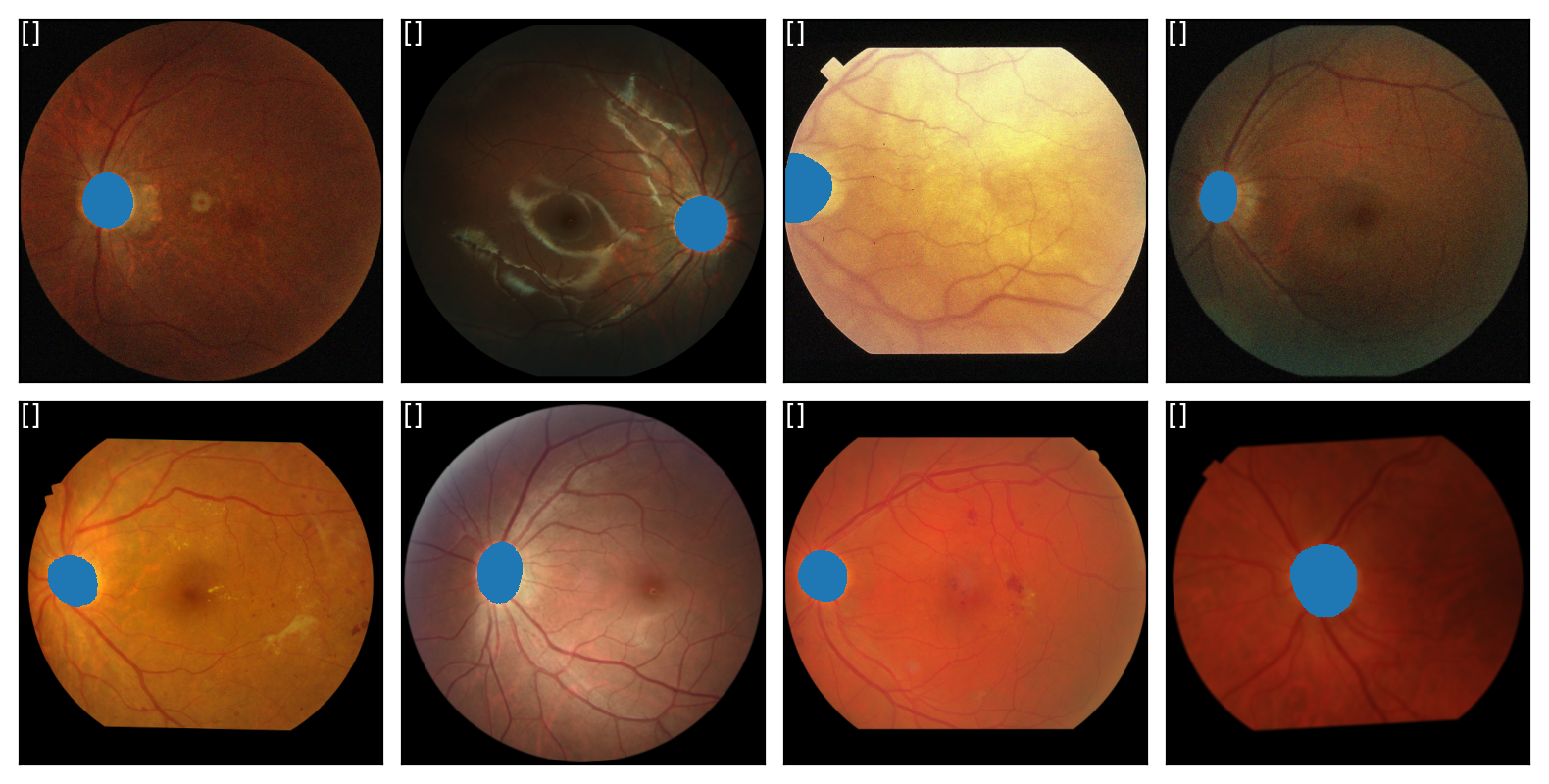}
        \caption{Disc segmentation masks.}
        \label{fig:batch_disc}
    \end{subfigure}
    \hfill
    \begin{subfigure}[b]{1.0\textwidth}
        \centering
        \includegraphics[width=\textwidth]{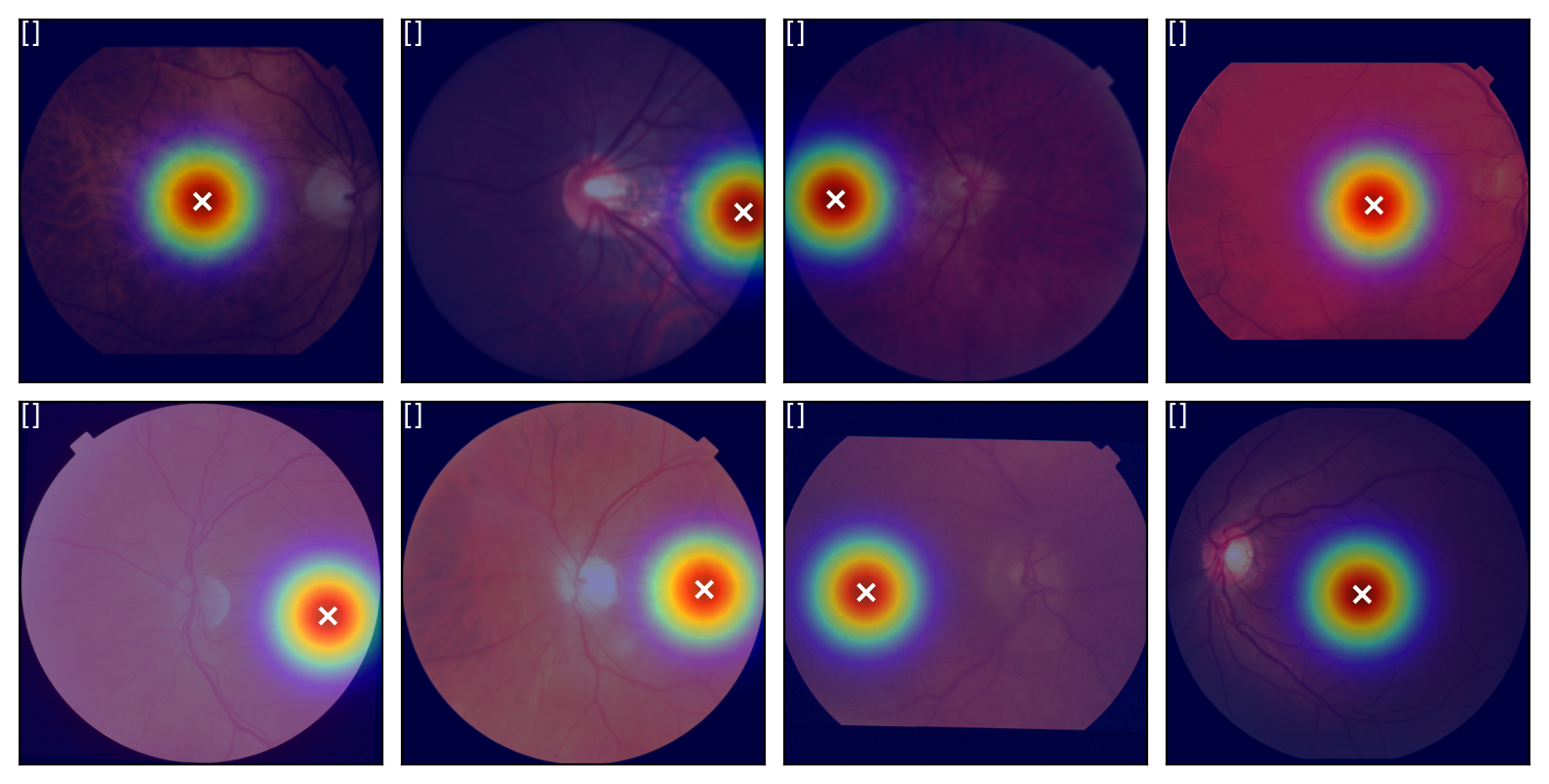}
        \caption{Fovea localization heatmaps overlaid over the input images.}
        \label{fig:batch_fovea}
    \end{subfigure}
    \caption{Sample training batches of input images to disc segmentation and fovea localization models after data augmentation. The contrast enhanced image is not shown.}
    \label{fig:batches_disc_fovea}
\end{figure}

\clearpage

\section{Feature Comparison}

\subsubsection{Feature implementations}\label{app:feature_details}

In this section we describe the implementation of our vascular features, or refer to previous work for a detailed description.

\begin{description}
    \item[Temporal Angle] We followed the implementation presented in \cite{gwas} (Section 2.3 of Supplementary Material). 
    \item[CRE] We followed the implementation presented in \cite{gwas} (Section 2.2 of Supplementary Material). However, we measure and aggregate all vessels on the temporal side of the image that intersect the defined circle. We use only the temporal side because the nasal side is often not visible at the evaluated radii in fovea-centered images.
    \item[Vascular Density]: Defined as the fraction of 1-pixels in the binary A/V segmentation mask within the fundus mask area:
    \[
        \text{Vascular Density} = \frac{\sum_{i=1}^{H} \sum_{j=1}^{W} I(i, j) \cdot M(i, j)}{\sum_{i=1}^{H} \sum_{j=1}^{W} M(i, j)}
    \]
    where:

    \begin{itemize}
        \item $I(i,j)$ is the binary vessel mask where each pixel is either 0 (background) or 1 (vessel).
        \item $M(i,j)$ is the binary mask where the region of interest (ROI) is defined, as extracted by our pre-processing algorithm
    \end{itemize}

    \item[Vessel Caliber] Vessel calibers or widths were measured per vessel segment by first fitting a spline to the segment skeleton. Perpendiculars to the spline were sampled at regular intervals. A ray casting method was used to extend the lines until they reached a background pixel (in the original artery or vein mask) on both sides, resulting in a width measurement. The median of the measurements was taken as the vessel segment caliber.

    \item[Tortuosity] Tortuosity was defined using its common definition as the length of the vessel following its points divided by its chord length (distance between its endpoints):

    \[
    T = \frac{\sum_{i=1}^{N-1} \sqrt{(x_{i+1} - x_i)^2 + (y_{i+1} - y_i)^2}}{\sqrt{(x_N - x_1)^2 + (y_N - y_1)^2}}
    \]

    where $(x_i, y_i)$ are the coordinates of the skeleton points, with $i = 1,2,...,N$ where $N$ is the total number of points along the vessel skeleton.

    \item[Curvature] Curvature is an alternative measure of tortuosity. Given a parametrization 
    $\gamma(t) = (x(t), y(t))$ of a twice differentiable plane curve, curvature can be expressed as:

    \[
    \kappa = \frac{|x' y'' - y' x''|}{(x'^2 + y'^2)^{3/2}}
    \]

    where the derivatives $x'$, $y'$, $x''$ and $y''$ are with respect to $t$.
    
    The mean curvature over the curve can be defined by integration of this expression:

     \[
    \bar{\kappa} = \int_0^1 \kappa(t)
    \]

    This calculation was approximated by computing $\kappa$ at regular intervals (every 5 pixels), using splines as parametrization to approximate derivatives.

    \item[Inflection count] Another alternative measure of tortuosity, this feature counts the number of inflection points (changes of direction) in a vessel segment. 

    \item[Num. Bifurcations] The number of bifurcations detected in the image. Bifurcations are detected by converting the skeleton into a graph, which is then traversed using the optic disc location as reference to defined outgoing vessels.

    \item[Bif. Angles] Smallest angle between the two child vessels in a bifurcation. A distance $\delta$ from the bifurcation point was defined, at which the angle is measured. The spline approximations are used to obtain points $(x,y)$ along the vessel segments at such $\delta$. Bifurcations with child vessels shorter than $\delta$ are ignored.

\end{description}

\begin{table}[!ht]
\centering
\begin{tabular}{rr|rrr|rrr}
\toprule
 & \multirow{2}{*}{Grader $\mu$} & \multicolumn{3}{c}{MAE} & \multicolumn{3}{c}{Pearson \textit{r}} \\
 & & VascX & AM & LWN & VascX & AM & LWN \\
\midrule
Temporal Angle - A & 127.49 & \textbf{3.44} & 5.40$\dagger$ & 10.48$\dagger$ & \textbf{0.81} & 0.72 & 0.56 \\
Temporal Angle - V & 132.46 & \textbf{3.02} & 4.85 & 11.68$\dagger$ & \textbf{0.86} & 0.74 & 0.39 \\
CRE - A & 12.28 & \textbf{1.81} & 1.98 & 2.98$\dagger$ & \textbf{0.68} & 0.38 & 0.22 \\
CRE - V & 19.98 & \textbf{1.98} & 2.80$\dagger$ & 4.70$\dagger$ & \textbf{0.74} & 0.58 & 0.14 \\
Vasc. Density - A & 4.76 & \textbf{0.58} & 1.15$\dagger$ & 1.42$\dagger$ & \textbf{0.77} & 0.61 & 0.48 \\
Vasc. Density - V & 5.43 & \textbf{0.58} & 1.01$\dagger$ & 1.31$\dagger$ & \textbf{0.79} & 0.63 & 0.39 \\
Vessel Caliber [med] - A & 5.25 & 1.37 & \textbf{1.33} & 1.73$\dagger$ & \textbf{0.20} & 0.18 & 0.09 \\
Vessel Caliber [std] - A & 3.06 & \textbf{0.43} & 0.65$\dagger$ & 0.65$\dagger$ & \textbf{0.73} & 0.57 & 0.47 \\
Vessel Caliber [med] - V & 5.36 & \textbf{1.50} & 1.58 & 2.24$\dagger$ & \textbf{0.45} & 0.36 & 0.15 \\
Vessel Caliber [std] - V & 4.81 & \textbf{0.49} & 0.89$\dagger$ & 0.74$\dagger$ & \textbf{0.76} & 0.67 & 0.36 \\
Tortuosity [med] - A & 1.09 & \textbf{0.0057} & 0.0064 & 0.008$\dagger$ & \textbf{0.82} & 0.78 & 0.71 \\
Tortuosity [med] - V & 1.08 & \textbf{0.0044} & 0.0046 & 0.0064$\dagger$ & \textbf{0.77} & 0.76 & 0.51 \\
Curvature [med] - A & 6.71 & \textbf{1.01} & 1.10 & 1.37$\dagger$ & \textbf{0.91} & 0.90 & 0.85 \\
Curvature [med] - V & 7.83 & \textbf{1.19} & 1.31 & 1.34 & \textbf{0.81} & 0.77 & 0.76 \\
Inflection count [med] - A & 1.69 & \textbf{0.51} & 0.56 & 0.69$\dagger$ & \textbf{0.38} & 0.19 & 0.22 \\
Inflection count [med] - V & 1.85 & \textbf{0.62} & 0.71 & 0.76* & \textbf{0.35} & 0.16 & 0.19 \\
Tortuosity [LW] - A & 1.12 & \textbf{0.023} & 0.027$\dagger$ & 0.06$\dagger$ & \textbf{0.17} & 0.15 & -0.29 \\
Tortuosity [LW] - V & 1.11 & \textbf{0.019} & 0.021$\dagger$ & 0.054$\dagger$ & \textbf{0.25} & 0.24 & 0.02 \\
Bif. Angles [mean] - A & 79.88 & 7.69 & \textbf{7.11} & 11.58$\dagger$ & \textbf{0.61} & 0.57 & 0.35 \\
Bif. Angles [med] - A & 78.46 & 7.83 & \textbf{7.24}* & 10.75* & \textbf{0.61} & 0.53 & 0.38 \\
Bif. Angles [mean] - V & 81.36 & 7.21 & \textbf{6.76} & 10.63$\dagger$ & \textbf{0.44} & 0.40 & 0.22 \\
Bif. Angles [med] - V & 81.55 & 6.77 & \textbf{6.56} & 9.51$\dagger$ & \textbf{0.50} & 0.42 & 0.26 \\
Num. Bifurcations - A & 15.97 & \textbf{6.43} & 6.84* & 7.62$\dagger$ & \textbf{0.70} & 0.70 & 0.68 \\
Num. Bifurcations - V & 18.11 & 7.79 & \textbf{7.64} & 9.11$\dagger$ & \textbf{0.76} & 0.71 & 0.61 \\
    \bottomrule
    \end{tabular}
    \caption{Feature quality comparison on the Leuven-Haifa dataset (240 CFIs). Results of comparing MAE of features extracted using different models for artery-vein segmentation: VascX, Automorph (AM) and LittleWNet (LWN). Features extracted from ground truth segmentations were used as the reference for MAE. For vessel or bifurcation features, the aggregation function (applied on features from the whole image) is indicated in brackes: [med]: median, [std]: standard deviation, [mean]: mean value, [LW]: length-weighted; vessels were weighted by their length. The feature implementation was the same across models. Grader $\mu$ is the mean of feature values; mean MAE is the mean absolute error between ground truth features and features extracted from model outputs. Pearson coefficients are correlations between ground truth and model outputs. *: Significant difference in MAE ($p < 0.05$) between the model and VascX. $\dagger$: $p < 0.001$}
    \label{tab:features_leuven}
\end{table}

\end{document}